# An Intelligent Software Workflow Process Design for Location Management on Mobile Devices

N. Mallikharjuna Rao

Associate Professor, Annamacharya PG College of Computer Studies

Newboyanapalli, RAJAMPET, Andhra Pradesh, INDIA, E-mail: drmallik2009@gmail.com , Seetharam.p@gmail.com

P.Seetharam

System Engineer, AITS

*Abstract-* **Advances in the technologies of networking, wireless communication and trimness of computers lead to the rapid development in mobile communication infrastructure, and have drastically changed information processing on mobile devices. Users carrying portable devices can freely move around, while still connected to the network. This provides flexibility in accessing information anywhere at any time. For improving more flexibility on mobile device, the new challenges in designing software systems for mobile networks include location and mobility management, channel allocation, power saving and security. In this paper, we are proposing intelligent software tool for software design on mobile devices to fulfill the new challenges on mobile location and mobility management. In this study, the proposed Business Process Redesign (BPR) concept is aims at an extension of the capabilities of an existing, widely used process modeling tool in industry with 'Intelligent' capabilities to suggest favorable alternatives to an existing software workflow design for improving flexibilities on mobile devices.**

## I. INTRODUCTION

Technology improvements authorize the building of information systems which can be used any place, any time, through mobile phones and wireless devices through networks. Mobile-wireless systems can create more benefits for organizations: e.g., productivity improvement processes and procedures flexibility, customer services improvement and information correctness for decision makers, which together stress competitive strategy, lower operation costs and improved business and subscribers processes.

*A. Mobile-Wireless Systems*

Schiller [2] describes two mobility extents: one is user mobility, which allows creating the connection to the system from different geographical sites. A Mobile-wireless network is a radio network distributed over land areas called cells, each served by at least one fixed-location transceiver known as a cell site or base station. When joined together these cells provide radio coverage over a wide geographic area. This enables a large number of portable transceivers (e.g., mobile phones) to communicate with each other and with fixed transceivers and telephones anywhere in the network, via base stations, even if some of the transceivers are moving through more than one cell during transmission and second one is device mobility, which enables mobility of both user and devices, such as mobile phones.

*B. Types of Mobile-Wireless System Applications*

Mobile-wireless systems are classified in to two types [3]: Horizontal application and vertical applications.

*1) Horizontal application*

This is a flexible to an extensive range of users and organizations for retrieving the data from the devices, e.g. e-mail, browsers and file transfer applications.

*2) Vertical Application*

Vertical applications are precise to a type of users and organization. For example: financial applications, such as money transfer, stock exchange and information enquiry; marketing and advertising applications according to the actual user positions, i.e., pushing coupons to stores and information about sales nearby; emergency applications to check real-time information from government and medical databases and utility companies applications used by technicians and meter readers. In view of quick developments and usage with mobile devices are required to improve the software design approach for vertical applications for better outcome.

*C. Problems in Mobile-Wireless Systems*

Mobile-Wireless systems face some exceptional problems originating from the mobile devices. First, these devices have small memories, short battery life, and limited calculation and computation capabilities. Second, there is a wide variety of devices, possessing different characteristics, and the application must be adaptable to all of them. Third, the use of the devices is tight because of their size, tiny screens, low resolution, and small keyboards that are difficult to operate. Fourth, security problems can arise when devices are lost, due to possible illegal access to sensitive data. Fifth, location identification and profiles retrieval from database of the subscribers while the subscribers are on roaming. There is another problem we found that, for probing the profiles from the Home Location register and transfer to the Visitor Location register while the subscriber are in roaming. It is increase the transition between the HLR and VLR databases in mobile-wireless networks.

This study uses software engineering to define the mobile-wireless systems quality components and develops an approach to quantify these components, in order to enable the evaluation, comparison, and analysis of mobile-wireless system quality. Mobile-wireless systems must be measured on the basis of traditional systems e.g. easiness of maintainability, minimum complexity, lack of faults, and mean time between system failures in mobile device. In view of above, for increasing flexibility on wireless systems, we are introducing the concept of intelligent software workflow design for mobile devices. In this paper, we are proposing software workflow for only location management strategy on mobile devices.

On the other hand, fuzzy logic can help a lot for in developing software for financial markets, environment control, project control and other scientific applications. It can help in detecting risks as early as possible in an uncertain environment. More important factor is that, it allows us to use common sense and knowledge for risk detection in software development. It provides an easy set of mathematical tools for assessing risks associated with software project management and software quality control within mobile devices for location based services. In section 2 we have discussed about Existing and its related work, section 3 we have presented proposed intelligent software workflow architecture, section 4 we have discussed about performance of the system and in section 5 we concluded the proposed system.

II RELATED WORK

The past fifty years of software engineering can be seen as a invariable hunt of developing software systems that are easier to design, cheaper to maintain, and more robust. Several milestones have been achieved in this effort, including data-oriented software architecture, object-oriented language, component-based software development and document-based interoperability. Among these achievements, in all the developments researchers have concentrated on two important concepts; they are data independence and process independence [4].

Data independence can be determined as the robustness of business applications when the data structures are modified in traditional software developments. Fuzzy database will become dominant in the database market in future, generally as it achieves significant role in data independence for better results in future.

Process independence is a measure of the robustness of business applications when the process model is redesigned. The drive towards more process independence has lead to the increase in the workflow of the systems in software industry in the last few years. Very recently, most major software companies have either acquired or developed workflow components for integration into their existing software platforms. Fuzzy database is giving more scope for getting many more decision making possibilities for redesigning the system with precise results.

Workflow supervision helps achieve software flexibility by modeling business processes unambiguously and managing business processes as data that are much easier to modify than conventional program modules. Workflow management systems enable reprocess of process templates, robust integration of enterprise applications, and flexible coordination of human agents and teams.

In present market, few Fuzzy Logic software tools are available to use in databases for solving complex problems faced by science and technology applications. They are presented in [5] they are: eGrabber DupChecker, Fuzzy System Component (Math Tools), Fuzzy Dupes 2007 5.6 Kroll Software-Development, Fuzzy Dupes Parallel Edition 32/64-Bit 6.0.3 (Kroll Software-Development), Sunshine Cards 1 (Free Spirits). In this paper, we took eGrabber DupChecker scenario for eliminating duplicate data from the database.

eGrabber DupChecker is a tool that uses advanced fuzzy logic technology to quickly identify hard-to-find duplicates in your Database. DupChecker automatically scans, identifies and groups duplicates for easy merging and de-duping.

*A. Features*

- *Robust Duplicate Checking*-Uses Fuzzy Logic Technology to quickly identify duplicates created through errors in data-entry, data-import and other variation such as Typo Errors, Pronunciation Errors, Abbreviation, Common nicknames, Match name with email ID, Company ending variations, Street ending variation, Formatting and punctuation errors, Field swap errors, Also recognizes same phone, regions and same locations.

- Merges all duplicates at one click on Merges notes, histories, activities, opportunities and all other details in Database.

- Retains history of all changes made to the merged record on databases

- Two plus merging - Merges multiple matches into one record

*B. Benefits*

- Maintains integrity of existing databases - Aggregates information lost across (notes, opportunities etc.,) duplicate contacts into one contact.

- Protects your company credibility - Avoid duplicated emails, postages or calls to customers or prospects, which would put your company creditability at stake.

- Saves time and money - Your sales team will not waste time calling duplicate contacts and you will cut down promotional cost spent for duplicate contacts.

Despite of recent significant developments in workflow technology and its prevalent acceptance in practice, current workflow management systems still demand significant costs in system design and implementation. Furthermore, these systems are still lacking in their ability of handling exceptions and dealing with changes. In this paper, we study additional techniques based on intelligent workflow to incorporate more flexibility in conventional workflow management systems. Our research goal is to further improvement in software flexibility by achieving more process independences using fuzzy logic.

### III PROPOSED METHOD

Business Process Redesign (BPR) is a popular methodology for companies to boost the performance of their operations. In core, it combines an essential reformation of a business process with a wide-scale application of Information Technology (IT). However, BPR on the work flow currently more closely resemble the performance of an art than of science. Available precise methods hardly find their way to practice.

The keywords for BPR are 'Fundamental', 'Radical', 'Dramatic', 'Change' and 'Process'. A business process has to undergo fundamental changes to improve productivity and quality. Radical changes, as opposite to incremental changes, are made to create dramatic improvements. Reengineering is not about fine-tuning or marginal changes. BPR is for provoked companies like mobile service providers; that are willing to make important changes to achieve significant performance improvements in their organization.

BPR is a structured approach for analyzing and continually improving fundamental activities such as manufacturing, marketing, communications and other major elements of a company's operation. Wright and Yu (1998) defined the set of factors to be measured before actual BPR starts and developed a model for identifying the tools for Business Process Redesign. The BPR is to develop a framework for understanding Business Process Re-engineering and to explain the relationship between BPR and Total Quality Management (TQM), Time-based competition (TBC) and Information Technology (IT). BPR should enable firms to model and analyze the processes that supports products and services, high-light opportunities for both radical and incremental business improvements through the identification and removal of waste and inefficiency, and implement improvements through a combination of Information Technology and good working practices in their work places.

*A. A framework for BPR modelling and analysis.*

The proposed framework has been presented for location based services on mobile devices, is offer some guidelines for choosing suitable tools/techniques for BPR applications. The guidelines are based on the areas to be reengineered for dramatic improvements in the performance.

*1) BPR Strategies*

Decision making at strategic levels would require intelligent systems to select the appropriate strategies and methods with the objective of making decisions about business location, product portfolio, funding for project work, etc. this requires taking into account the risk involved and the costs and benefits of running the business. At strategic levels, aggregate and fuzzy data are used to make a decision for long-term developments and changes in an organization. The type of decisions requires experience and knowledge in selecting the most suitable methods.

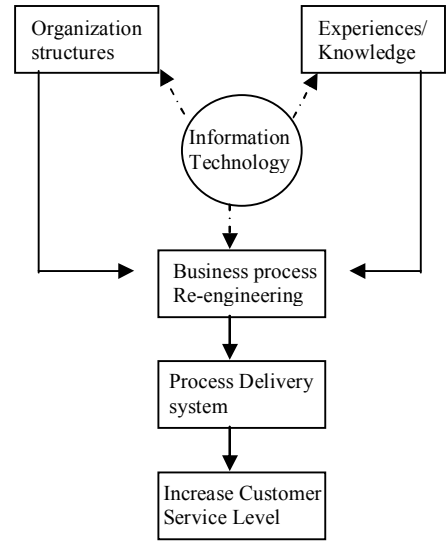

Figure 1: A Conceptual Intelligent Business model

A Conceptual Intelligent Business Model as shown in figure 1; which contains organization structures, knowledge data interact with information technology for more and instant decisions to redesign and increasing the outcome of the devices. Therefore, the selection of the tool for BPR depends upon:

1) The nature of decision areas.

2) The nature of data to be analyzed.

3) The background of users.

The decision area here is to formulate strategies for reengineering business processes. The nature of data available at the strategic level is generally not accurate at this level of decision making, and therefore models based on a system approach and conceptual framework could be used to analyze the data. Knowledge-based models are user-friendly, but have limited applications considering the areas of reengineering.

In this study, the proposed Intelligent Software Workflow process design as shown in figure 2; consists of several blocks. In this system, all the system protocols are connected to the set of attributes to software design they can automatically reflect the design process on devices and device performance. As I, defined already, fuzzy logic is an intelligent system which can produce precise data values; they will give ultimate outcome for the devices. Fuzzy logic is a rule based system, with number of rules it can generate more and more combination of results which can help to re-construct or re resign based on the values produced by the system.

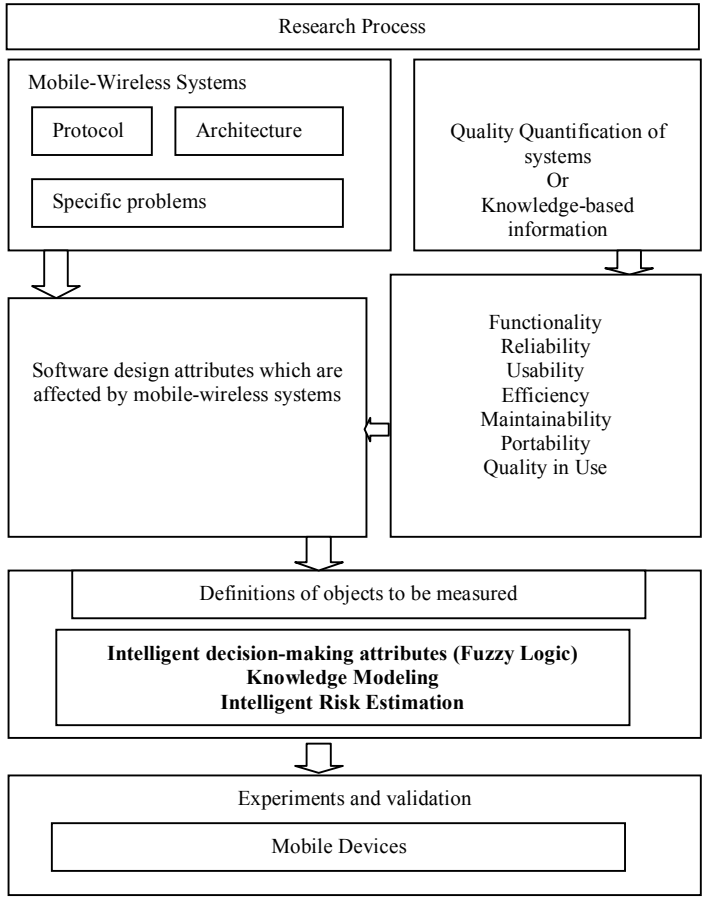

Figure 2: Intelligent Software Workflow Process Design Architecture

This can increase the functionality of the system; it includes suitability for found the call, accuracy on making the calls, interoperability and security. Increase the Reliability which includes; maturity, fault tolerance and recoverability. During mobility, network problems, hiding obstacles and hopping between antennas may disturb and interrupt communications.

In Usability, this includes the understandability, learnability, operability and attractiveness. Mobile users may not be able to focus on the systems use, so the application should not be complicated, the input must be easy to insert, intuitive and simplified by using location aware functions. Efficiency includes the time behavior and resource utilization. Time behavior in the wireless environment because the price of each minute of data transferring is very high, and the users will avoid expensive systems.

Maintainability includes the analyzability, changeability, stability and testability. Portability includes the adaptability, installability, co-existence and replacebility.

The following resources can used in mobile devices to design the workflow process for location management. They are: Mobile Network Code (MNC), Location Area Code (LAC), Base Station Identity (CID) and Radio Frequency (RF) Signal– The relevant signal strength between your current tower and the adjacent one.

In this study, we took above attributes as vague input values and processed through the proposed workflow system, it make many more decisions and knowledge decisions and the outcome of this system produces well behaved work and timeless connection and transfer of text messages from the one terminal to another terminal without interruptions. It will increase the performance, stability and operational flexibility on mobile devices while the subscribers are in roaming.

In the next section, we have discussed few important aspects to design best and better software workflow design using fuzzy logic which enables flexibility on usage and decrease the transmission delays.

*B. Knowledge Modeling*

Knowledge is information combined with aptitude, framework, investigation, and suggestion. It is a high-value form of information that is ready to apply to decisions and actions. Knowledge Modeling packages combinations of data or information into a reusable format for the purpose of preserving, improving, and allocation, aggregate and processing Knowledge to simulate intelligence. In intelligent systems quality of the knowledge base and usability of the system can be preserved by filtering the domain expert is involved in the process of knowledge acquisition but also in other phases until the testing phase. The quality and usability is also dependent of the end users consideration why these users would be involved in the modeling process and until the system has been delivered.

Meta-rules are rules that are combining ground rules with relationships or other meta-rules. Thereby, the meta-rules tightly connect the other rules in the knowledge base, which becomes more consistent and more closely connected with the use of meta-rules.

Vagueness is a part of the domain knowledge and refers to the degree of plausibility of the statements. This is easy to get outcome from something in between, e.g. unlikely, probably, rather likely or possible, the rule based representation shown in the figure 3.

Knowledge base contains a number of fuzzy linguistic variables (fuzzy sets), and a number of fuzzy inference rules. Semantics fuzzy linguistic variable (fuzzy sets) are defined by their membership functions based on software metrics. It contains a list of fuzzy inference rules about risk detection across all phases in software life cycles. Fuzzy rules are basically of "IF-THEN" structure.

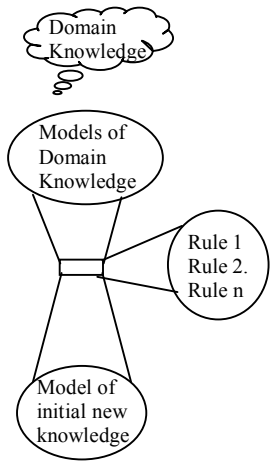

Figure 3: Knowledge modeling

Fuzzy inference rules are presented in antecedent-consequences structure. Antecedents represent symptoms of software artifacts, processes or organization in terms of risks based on the 'Dimensional Analytical Model' rules use individual metrics and their combinations based on their relational ships from different dimensions.

*C. Intelligent Risk Estimation*

Intelligent risk estimation is the result of fuzzy inference engine as shown in figure 4, which is the central processing part of the intelligent software workflow system. In this system all the metrics have certain values and it also have the three dimensions. Three dimensions are connected to Inference Engine for making more decisions. Inference engine apply the knowledge base on this set of inputs to produce worth risk, which is called schedule low risk, standard risks and high risk. The input and output sets are stored over the time period for analysis purpose [7]. This analysis helps the mobile location management software designers to find the maximum predictable risk for the systems, even network failures or minimum number of cell towers or even low frequency. This system can produced LR (Low Risk), SR (Standard Risk) and HR (High Risk).

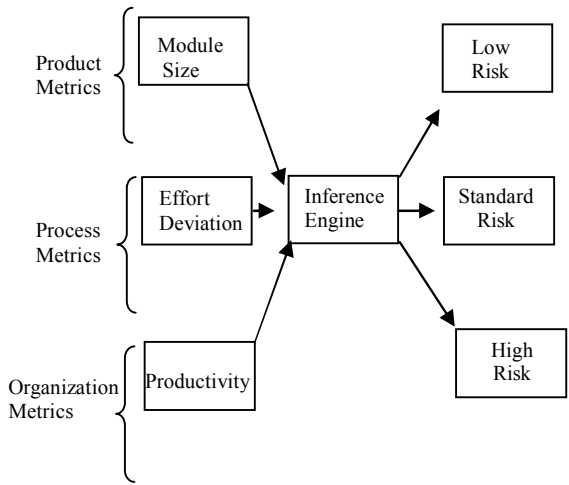

Figure 4: Fuzzy Inference Engine for Intelligent Risk Assessment

In this study, for finding risk estimation for location identification on mobile devices, measures the number of mobiles connected to networks, networks size and all other protocols are required to include into the system for finding the risk of the system. Fuzzy inference engine is evaluating the risk factor based set of inputs, Module size, effort deviation and Productivity. Inference engine take the inputs of the systems and it produces the out of low risk, standard risk and high risk factors. We are given some set of conditions rules for improving the system. They are

*1) Product Metric based rule*

IF Volatility index of subsystem is HIGH
AND Requirements quality is LOW
THEN Schedule Risk is VERY HIGH

*2) Process Metric based rule*

IF Manpower is HIGH
AND Design approaches are HIGH
THEN Product Service is HIGH

*3) Organization Metric based Rule*

IF Effort deviation is HIGH
AND Customer involvement is HIGH
THEN Risk of schedule is VERY HIGH

IV SIMULATIONS ON PROPOSED SYSTEM

In this study, intelligent software workflow system has been developed for fuzzy query, which makes it possible to query on a conventional databases. The developed software fuzzifies the fields, which are required to be fuzzified, by connecting to the classical database and creates a supplementary database. This database includes fuzzy data and values which are related with the created membership functions. In the recent years, fuzzy query usage is increase in all the areas in the intelligent systems.

A database, called as "Subscriber profile", which includes some information about the mobile subscribers. Suppose that a query for the approximation of the necessity of the subscriber's profiles is made. In order to determine details of the Subscribers in normal database is difficult. For example, the SQL query for the particular assessment problem may be in the form of the following.

SELECT subscriber_name, imei#, sim#, La, mobile#, bill_payment FROM *SUBSCRIBER_PROFILE* WHERE bill_payment <=3000

Same query convert into fuzzy SQL is as follows:

SELECT subscriber_name, imei#, sim#, La, mobile#, bill_payment FROM *SUBSCRIBER_PROFILE* WHERE bill_payment is HIGH or more than 3000

That means we will get the subscribers those who can the have the more than 3000 payment or payment HIGH subscribers.

*A. FUZZY QUERY SOFTWARE TOOL*

It is intrinsically robust since it does not require precise, noise-free inputs and can be programmed to fail safely if a feedback sensor quits or is destroyed. The output control is a smooth organize function despite a wide range of input variations. The membership function is a pictographic representation of the degree of participation of each input. It associates a weighting with each of the inputs that are processed, define functional overlap between inputs, and ultimately determines an output response. Once the functions are indirect, scaled, and combined, they are defuzzified into a crisp output which drives the system.

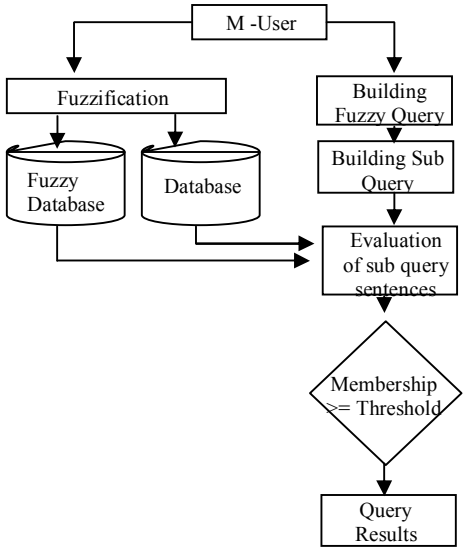

Figure 5: The Block diagram of the fuzzy tool

As it is known, a fuzzy set is a set, which contains elements having varying degrees of membership. Elements of a fuzzy set are mapped into a universe of "membership values" using a function-theoretic form. Fuzzy sets are denoted by a set symbol. As an example, let "A" denote a fuzzy set. Function $\mu$ maps the elements of fuzzy set "A" to real numbered value on the interval 0 to 1. If an element in the universe, say x, is member of the fuzzy set "A", then this mapping is given as:

$$\mu_A(x) \in [0,1]$$
$$A = (x, \mu_A(x) / x \in X)$$
$$A = \sum \mu_A(x_i) / x_i = \mu_A(x_1) + \mu_A(x_2) + ..... + \mu_A(x_n) / x_n$$

Using above equation, we can calculate the membership degrees for crisp variables in a system.

In this study, we presented fuzzy representation for the risk estimation using rule based crisp values and membership degrees as we shown in figure 6 and figure 7.

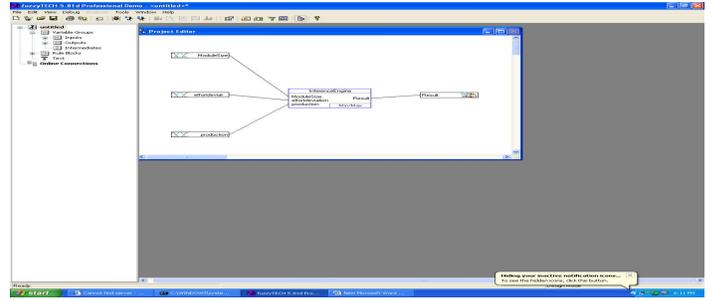

Figure 6: Fuzzy variable editor making fuzzy decisions

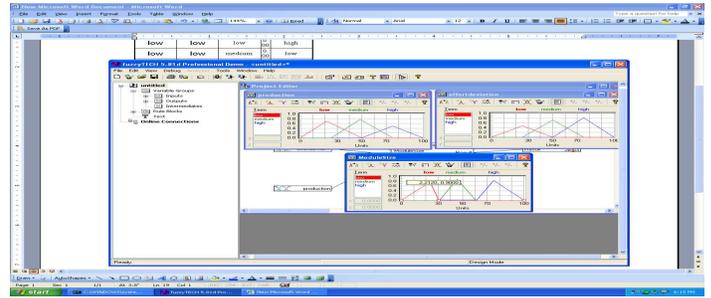

Figure 7: Membership representation

## CONCLUSION

This paper introduces new concept called intelligent software design. First, it describes mobile-wireless systems and stated the new challenges faced by mobiles devices on software modeling. Second, this study gives better solution for location and mobility management software design is improve the flexibility to transfer and processing the information from one location to another location. This Intelligent software design tool can be expanded to new kinds of mobile-wireless systems, emerging because of the rapid development of the technology and the wireless networks.